\documentclass[12pt,twoside]{article}
\usepackage{psfig}
\newcommand{\VolumeHeader}{}
\newcommand{\VolumeSerial}{LNS}

\newcommand{\ActivityName}{ {\normalsize {\it 
Conference on Gravitational Waves:
}}\\
{\normalsize {\it 
 A Challenge for Theoretical Astrophysics
}}}
\newcommand{\ActivityDate}{ {\normalsize {\it
Trieste,  5-9 June (2000) 
}}}

\newcommand{\be}{\begin{equation}}
\newcommand{\ee}{\end{equation}}
\newcommand{\bea}{\begin{eqnarray}}
\newcommand{\eea}{\end{eqnarray}}

\newcommand{\Fstar}{ \raisebox{.2ex}{$\stackrel{*}{F}${} } }

\newcommand{\LectureHeader}{Astrophysical tests for the NDL theory of gravity}

\begin{document}
\pagestyle{myheadings}
\markboth{\LectureHeader}{\VolumeHeader}
\markright{\VolumeHeader}


\begin{titlepage}


\title{Astrophysical Tests for the Novello-De Lorenci-Luciane Theory of Gravity: \\ Gravitational-Wave + Neutrino Bursts from Local Supernovae and Gravitational-Wave Microlensing by Galactic (MACHOs) Black Holes} 

\author{Herman J. Mosquera Cuesta$^{1,2}$\thanks{herman@ictp.trieste.it :: herman@lafex.cbpf.br}
\\[1cm]
{\normalsize
{\it $^1$Abdus Salam International Centre for Theoretical Physics}}\\
{\normalsize
{\it Strada Costiera 11, Miramare 34014, Trieste, Italy. }} \\
{\normalsize
{\it $^2$Centro  Brasileiro  de Pesquisas  F\'{\i}sicas}} \\ {\normalsize
{\it Laborat\'orio  de Cosmologia e F\'{\i}sica  Experimental  de Altas Energias}}\\
{\normalsize
{\it Rua Dr.  Xavier Sigaud 150, Cep 22290-180, Rio de Janeiro, RJ, Brazil.}}\\ 
\\[10cm]
{\normalsize {\it Lecture given at the: }}
\\
\ActivityName 
\\
\ActivityDate 
\\[1cm]
{\small \VolumeSerial} 
}
\date{}
\maketitle
\thispagestyle{empty}
\end{titlepage}

\baselineskip=14pt
\newpage
\thispagestyle{empty}


\begin{abstract}

The Novello-DeLorenci-Luciane (NDL) field theory of gravitation predicts that gravitational waves (GWs) follow geodesics of a modified (effective) geometry with a speed lower than the velocity of light. The theory also demonstrates that GWs exhibit the phenomenon of birefringence, formerly believed to be exclusive of electromagnetic waves. Here prospective astrophysical tests of these predictions are proposed. I point out that future measurements of gravitational waves in coincidence with a non-gravitational process such as {\it a neutrino burst} (and likely a burst of gamma-rays) may prove useful to discriminate among all the existing theories of gravity. It is also stressed that microlensing of gravitational waves emitted by known galactic sources (i.e., pulsars) in the bulge, lensed by either the Galaxy's central black hole (Sgr A$^\ast$) or a MACHO object adrift among the Milky Way's stars, may provide a clean test of the birefringence phenomenon implied by the NDL gravity theory.

\end{abstract}

\vspace{6cm}




\newpage
\thispagestyle{empty}
\tableofcontents

\newpage
\setcounter{page}{1}

\section{Einstein's Theory and the Equivalence Principle}

A succesful theory of gravity should be able to correctly predict the way this interaction occurs in all process in nature. Einstein's theory of gravitation has till now passed all of the tests in this concern. However, it encompasses an implicit statement concerning the way gravity-gravity interaction develops when compared to gravity-nongravitational energy interactions. General relativity stands on the {\it equivalence principle}, which states that any sort of matter including massless fields like the photon, interacts with gravitational fields fundamentally inasmuch as the same manner. This statement allows to interpret all the gravitational interactions, including gravity-gravity as well (this one having no any experimental or observational foundation), as due to changes in the space-time geometry induced by the presence of matter fields: $g_{\mu \nu} =  \gamma_{\mu \nu} + \varphi_{\mu \nu}$. However, if one dismisses the assumption that the gravitational energy should encompass the hypothesis of universality of the equivalence principle, i. e., Einstein equivalence principle does not apply to free falling "gravitons'', a field theory of gravity in which the gravity-gravity interaction occurs in a rather different way compared to gravity-nongravity can be formulated\cite{NDL}.

The Novello-DeLorenci-Luciane (NDL) theory of gravitation has been recently introduced\cite{NDL}. It was shown it incorporates essentially all the ingredients general relativity endowes\cite{NDL}, and in this vein it resembles Einstein theory as far as the first post-Newtonian approximation for solar system tests is concerned, and also for the radiative solution up to the quadrupole formula level. It has been demonstrated that the most striking prediction of the NDL theory is related to the velocity of propagation of gravitational perturbations\cite{PLA99}\footnote{A new more stringent test of the NDL theory predictions concerning the birefringence of the GWs will be addressed in section 6\cite{MVH00}. It is shown there that birefringence of GWs is a peculiar characteristic of almost all non-linear theories of gravity, {\it except general relativity}, and in particular of the NDL.}. In Ref.\cite{PLA99} was shown that gravitational waves (GWs) travels in the null cone of an effective geometry with a speed lower than the velocity of light, the one for GWs to travel in Einstein's theory.


\section{The Novello-DeLorenci-Luciane (NDL) Field Theory of Gravity}

In a previous paper \cite{NDL} a modification of the
standard Feynman-Deser approach of field theoretical derivation of Einstein\rq s general relativity, which led to a competitive 
gravitational theory, was presented. 
The main lines of such NDL approach can be summarized as follows:
\begin{itemize}
 \item{Gravity is described by a symmetric second rank tensor 
$\varphi_{\mu\nu}$ that satisfies a non-linear equation of motion;}
 \item{Matter couples to gravity in an universal way. In this interaction, 
the gravitational field appears only in the combination 
$ \gamma_{\mu\nu} + \varphi_{\mu\nu}$, inducing us to define a quantity
$ g_{\mu\nu} = \gamma_{\mu\nu} + \varphi_{\mu\nu}$. This tensor 
$g_{\mu\nu}$ acts as an effective metric tensor of the 
spacetime as seen by matter 
or energy of any form except gravitational energy;}
 \item{The self-interaction of the gravitational field breaks the
above universal modification of the spacetime geometry.}
\end{itemize}

\subsection{Notation and Definitions.}

We define a three-index tensor $F_{\alpha\beta\mu}$, which we will call 
the {\bf gravitational field}, in terms of the symmetric standard variable 
$\varphi_{\mu\nu}$ (which will be treated as the potential) to describe a
spin-two field, by the expression\footnote{We are using the 
anti-symmetrization symbol $[x, y] \equiv xy - yx$ and the symmetrization symbol  $(x, y) \equiv xy + yx$. Note that indices are raised and lowered by the Minskowski background metric $\gamma_{\mu\nu}$. The covariant
derivative is denoted by a semicomma `$;$' and it is constructed 
with this metric.} $F_{\alpha\beta\mu} = \frac{1}{2} ( \varphi_{\mu[\alpha;\beta]} + 
F_{[\alpha}\gamma_{\beta]\mu} )
\label{d1}$, where $F_{\alpha}$ represents the trace: $F_{\alpha} = F_{\alpha\mu\nu} \gamma^{\mu\nu} = \varphi_{,\alpha} - 
\varphi_{\alpha\mu;\nu} \gamma^{\mu\nu}$.

From the above definition it follows that $F_{\alpha\beta\mu}$ is 
anti-symmetric in the first pair of indices and obeys the cyclic 
identity, that is: $F_{\alpha\mu\nu} + F_{\mu\alpha\nu} = 0
\label{d2}$ and $F_{\alpha\mu\nu} + F_{\mu\nu\alpha} + F_{\nu\alpha\mu} = 0.\label{d3}$

The most general non-linear theory must be a function 
of the invariants one can construct with the field.
There are three of them which we represent by $M$,$ N$ and $W$, that is: $M \equiv F_{\alpha\mu\nu}\hspace{0.5mm} F^{\alpha\mu\nu},\nonumber
N \equiv F_{\mu}\hspace{0.5mm} F^{\mu}, W \equiv  F_{\alpha\beta\lambda} {\Fstar}^{\alpha\beta\lambda} \doteq \frac{1}{2} F_{\alpha\beta\lambda}\hspace{0.5mm} F^{\mu\nu\lambda}\,
\eta^{\alpha\beta}\mbox{}_{\mu\nu}.\label{AB}$

We will deal here only with the two invariants $U \equiv M - N$ and
$W$. The reason for this rests on the linear limit. Indeed, 
in order to obtain the standard Fierz linear theory ---as it is required of any
candidate to represents the dynamics of spin-two--- the invariants
$M$ and $N$ should appear only in the combination $U$. This is the case,
for instance in Einstein General Relativity theory.

Under this condition, the general form of the Lagrangian density 
is given by: $L = L(U, W),\label{Lag}$ with the gravitational action expressed as: $S = \int{\rm d}^{4}x\sqrt{-\gamma}\,L,
\label{action}$  where $\gamma$ is the determinant of the flat spacetime  metric $\gamma_{\mu\nu}$ written in an arbitrary coordinate system.

From the Hamilton principle we find the following equation of motion 
in the absence of material sources:

\begin{equation}
\left[L_{U} F^{\lambda (\mu\nu)} + L_{W}\, {\Fstar}^{\lambda (\mu\nu)} 
\right]_{;\lambda} = 0.
\label{eqmov}
\end{equation}

$L_{X}$ represents the derivative of the Lagrangian with respect to the invariant $X,$ which may be $U$ or $W$.


\section{Velocity of Gravitational Waves}

The GWs dispersion relation in the NDL theory reads: $k_\mu k_\nu[\gamma^{\mu \nu} + \Lambda^{\mu \nu}] = 0,$ where $\Lambda^{\mu \nu} = 2 \frac{L_{UU}}{L_U}[ F^{\mu \nu \beta} F^\nu_{(\alpha\beta)} -  F^\mu F^\nu],\label{3-tensor}$ with $L_U$ and $L_{UU}$ corresponding,  respectively, to the first and  second derivative of the Lagrangian of the theory respect to the invariant $U$, defined below. 

Thence the discontinuities of the gravitational fields propagate in a modified geometry which changes the background geometry $\gamma^{\mu \nu}$ (the Minkowski metric) into an effective one

\be
g^{\mu \nu}_{eff} \equiv \gamma^{\mu \nu} + \Lambda^{\mu \nu},
\ee

which has dependence upon the field $F_{\alpha \beta \mu}$ and its dynamics. The overall characteristic of the new geometry is determined by the  non-linear character of the lagrangian on which the theory is based. Then the GWs velocity (for a massless graviton) in the NDL reads

\be
v_k^2  =  1 - \frac{1}{2b^2} \frac{1}{[1 + (k/b^2) {\cal{L}}]^2} Z^{\mu \nu} \frac{k_\mu}{|\vec{k}|}  \frac{k_\nu}{|\vec{k}|},\label{speed}
\ee

with the velocity of light $c = 1$ in geometric units. Here we define $Z^{\mu \nu} = F^{\mu (\alpha \beta)} F^\nu_{(\alpha \beta)} -  F^\mu F^\nu$. In the expression for the velocity of the GWs, Eq.(\ref{speed}), the Born-Infeld type Lagrangian density 

\be
{\cal{L}} = \frac{b^2}{k} \left[\sqrt{ 1 - \frac{U}{b^2} } - 1\right],
\ee

with $b$ a constant and $ k \equiv \frac{8\pi G_N}{c^4}$, is the most general functional of the invariant of the theory $U$. The quantity $U$, the dynamical parameter of the NDL theory, is defined in terms of the two fundamental invariants of the theory: M and N. Note that in the linear regime ${\cal L}(U) = U$. We then obtain the standard weak-field limit as it should be for any massless spin-2 theory of gravity, including general relativity.  The reader can see Ref.\cite{PLA99} for a more detailed discussion of the NDL gravitation.


Thus a crucial test of the NDL theory, and consequently a potential discriminator among the existing theories of gravity, could be an exact determination of the velocity of propagation of the GWs  themselves. This is an issue which is expected to be accomplished with the advent of the new generation of GW detectors such as the interferometers LIGO, VIRGO, GEO-600, and the TIGAs resonant-mass omni-directional observatories\cite{thorne95}. Below we suggest a prospective astrophysical experimental test of the NDL theory involving the detection of GWs in coincidence with a neutrino burst from a supernova explosion, including collapsars or hypernovae events.

We stress that the future detection of the GWs themselves (at least for one detector) is unable to provide the looked for discriminator criteria to settle this issue in the light of Einstein's and NDL theories. Therefore, a non-gravitational astrophysical or cosmological process is called for, and the expected neutrino bursts from both the {\it deleptonization} process in the supernova core and the gamma-ray burst surge accompanying the GWs in a hypernova event may prove useful.

\section{Neutrino-Driven Supernovae and Gravitational-Waves}

\subsection{Core-Collapse Neutrino-Driven Explosions}

During the precedent three decades most researchers in supernovae physics have explained type-II events as a consequence of neutrinos carrying the huge binding energy of the newly born neutron star. Then neutrinos deposit a portion of their energy in a low density region surrounding the star's core and a fireball of pairs and radiation finally explodes the remainings of the star. In these lines, core-collapse  supernovae  explosions are one of the  most 
powerful sources of neutrinos  $\nu_e, \nu_\mu,  \nu_\tau$ and its  antiparticles, and likely the sterile one $\nu_s$.  Different  theoretical and numerical  models of type II supernovae explosions \cite{woosley,janka96,janka97} have estimated that

\be  
\Delta  E_{total} =  5.2 \times 10^{53} {\rm erg}  \left(\frac{10\;km}{R_{NS}}\right) \left(\frac{M_{NS}}{ 1.4 \; M_\odot}\right)^2 \ee

are carried away by neutrinos.  Almost $\sim 10^{58}$  neutrinos of
mean  energies  $(10-25)$ MeV  are  released  over a time  scale of  seconds through the process $\gamma + \gamma \longrightarrow e^+ + e^- \longrightarrow \bar{\nu} + \nu$. Investigations  have shown that nearly 99\% of the total  gravitational  binding
energy of the  protoneutron  star can directly be carried away by these  neutrinos on their diffusion timescale $\Delta t_{nu} \sim 12$ s after the core bounce  $\Delta t_{CB} \sim 1$ ms \cite{woosley,janka96}. The remaining energy being radiated in electromagnetic and gravitational waves.

\subsection{Gravitational-Wave Characteristics from Local Supernovae}

On the other hand, during the core-collapse of supernova the  time-varying   anisotropic   distribution  of  density gradients in the proto-neutron star translates into the equivalent of a changing   quadrupole  mass-tensor  whose   dynamics   induces  emission  of gravitational   wave bursts\cite{burrows95}. Because the NDL theory agrees with general relativity upto the first post-Newtonian order, we can compute the amplitude of the GW signal as

\be
 h_{ij} = \frac{2 G}{c^4 D} \frac{d^2Q_{ij}}{dt^2} \longrightarrow h \sim  10^{(\{-18\} \{-19\})}
\ee

for distances as far as the Large Magellanic Cloud $D \sim 55$ kpc. Here $Q_{ij}$ defines the mass quadrupole tensor. This signal can last for hundred of milliseconds, with maximum GW frequency $\sim $ 1kHz. Since the GWs do not couple to any other form of energy they stream away from the SN core whereas  ordinary neutrinos in principle do not. This interaction induces a time-delay in the neutrino propagation respect to light, or equivalently to GWs in the Einstein theory of gravitation. We suggest that such time lag can be used also to test the prediction of the NDL theory that GWs travel at a speed lower than the corresponding one for light.

\subsection{Neutrinos from GRBs}

Current models of GRBs predict both ultra high, very high\cite{waxman} and high energy neutrinos\cite{meszaros} and ultra high energy cosmic rays emissions\cite{waxman} which may account for the extra-galactic high energy proton flux observed. Next we discuss how the most energetic neutrinos (expected to accompany the GWs burst from a collapsar) are emitted according to the GRBs standard fireball model. The reader can see Ref.\cite{waxman} for a more complete review of this mechanism. In the GRBs fireball picture the detected $\gamma$-rays are produced via synchroton radiation of ultrarelativistic electrons boosted by internal shocks of an expanding relativistic blast wave (wind) of electron-positron pairs, some baryons and a huge number of photons. The typical synchroton frequency is constrained by the characteristic energy of the accelerated electrons and also by the intensity of magnetic field in the emitting region. Since the electron synchroton cooling time is short compared to the wind expansion time, electrons lose their energy radiatively. The standard energy of the observed synchroton photons is given by

\be
E^b_\gamma = \frac{\Gamma \hbar \gamma^2_e e B}{m_e c} \simeq 4 \xi^{1/2}_B \xi^{3/2}_e \left(\frac{ L^{1/2}_{\gamma,51} } {\Gamma^2_{300} \Delta t_{\rm ms} } \right) {\rm MeV},
\ee

where $L_{\gamma,51}$ defines the energy released in GRBs with $L_\gamma = 10^{51} L_{\gamma,51}$ ergs$^{-1}$ the standard luminosity of BATSE observed GRBs, $\Delta t = 1 \Delta t_{\rm ms}$ ms is the typical timescale of variability, $\Gamma = 300\Gamma_{300}$ the Lorentz expansion factor, and $\xi_B$ corresponds to the fraction of energy carried by the magnetic field: $4 \pi r^2_d c \Gamma^2 B^2 = 8\pi \xi_B L$, being $L$ the total wind luminosity, and $\xi_e$ the one electrons carry away. No theory is available to provide specific values for both  $\xi_B$ and $\xi_e$. However, for values near the equipartition the model photons' break energy $E^b_\nu$ is in agreement with the observed one for $\Gamma \sim 300$ and $\Delta t = 1$ ms, as discussed below.

In the acceleration region protons (the fireball baryon load) are also expected to be shocked. Then their {\it photo-meson} interaction with observed burst photons should produce a surge of neutrinos almost simultaneously with the GRBs via the decay $\pi^+ \leftrightarrow \mu^+ + \nu_\mu \leftrightarrow e^+ + \nu_e + \bar{\nu}_\mu + \nu_\mu$. The neutrino spectrum in the fireball driven explosion follows the observed $\gamma$-rays spectrum, which approximates the broken power-law: $\frac{dN_\gamma}{dE_\gamma} \propto E^\beta_\gamma$, with $\beta \sim 1$ for low energies and $\beta \sim 2$ for high energies compared to the observed {\it break energy} $ E^\beta_\gamma \sim 1$ MeV, where $\beta$ changes. The interaction of protons accelerated to a power-law distribution: $\frac{dN_p}{dE_p} \propto E^{-2}_p$, with the fireball photons results in a broken power-law neutrino spectrum: $\frac{dN_\nu}{dE_\nu} \propto E^{-\beta}_\nu$, with $\beta =1$ for $E_\nu < E^b_\nu$, and $\beta = 2$ for $E_\nu > E^b_\nu$. Thus the neutrino break energy  $E^b_\nu$  is fixed by the threshold energy of photons for {\it photo-production} interacting with the dominant $\sim 1$ MeV fireball photons, and reads

\be
E^b_\nu \simeq 5 \times 10^{14} \Gamma^2_{300} \left(\frac{E^b_\gamma}{ 1 {\rm MeV}}\right)^{-1} {\rm eV}.
\ee


\section{Collapsars, Neutrino and Gravitational-Wave Bursts: A Test for the NDL Velocity of GWs}

The just described picture for driving supernovae explosions is by now being considered unable to explain the observational fact that some supernovae appear to require more energy (an order of magnitude higher) than is provided by the current mechanism based on neutrino transport\cite{woosley99}. Moreover, the trend in gamma-ray burst (GRBs) modelers is converging on a scenario in which a massive presupernova star (and its final explosion as a ''hypernova'') is the leading candidate\cite{woosley99}. This new paradigm {\it the collapsars}: supernovae explosions in which a stellar mass black hole, formed previously to the star final disruption, is the central engine for the GRBs. This model is supported by the fact that some supernovae have been found to be associated with GRBs events. The abrupt fallback ($\Delta T_{acc} \leq 10^{-3}$s)\footnote{This timescale will define also the main characteristic frequency of the GW signal emitted.} of a surrounding accretion disk, remnant of the failed supernova previous stage, triggers the emission of strong GRBs most likely accompanied by GWs and neutrino bursts. In our view, this model comprises the necessary non-gravitational astrophysical processes ($\gamma$ + $\nu$ bursts) through which we can stringently test the NDL theory concerning the velocity of propagation of GWs. For more details on the collapsar mechanism we address the reader to Ref.\cite{woosley99}, and references therein.

Thus, let us assume for a while that the gravitational radiation (including the GW burst produced during the neutrino outburst\cite{burrows95}) travels at the speed of light. Because of the observational evidence that neutrinos actually oscillate\cite{fukuda98}, which implies they endow a mass; and consequently cannot travel at the speed of light, we can use the analogous expression for computing the neutrino time delay compared to photons emanating from the heavy neutrino radiative decay channel, to estimate their proper time delay with respect to the gravitational radiation surge generated at core bounce.  Then the time delay for the neutrinos (emitted simultaneously with the burst of GWs) to arrive to the neutrino telescope is expressed as \cite{raffelt99,mohapatra,klapdor}

\be
\Delta T_{GWs \leftrightarrow \nu_s} = 0.515 \; {\rm s} \left(\frac{D}{10\; kpc}\right)  \left(\frac{m{^2_\nu}}{100\; eV^2}\right)  \left(\frac{100\; MeV^2}{E{^2_\nu}}\right),\label{neut}
\ee

where $E_\nu$ represents the neutrino energy, $D$ the source distance to Earth and $m_\nu$ the neutrino mass. Since there is a network (SNEWS\footnote{The SuperNova Early Warning System.}) of neutrino detectors currently running that are sensitive to the prompt core-collapse supernova neutrino bursts in our galaxy\cite{scholberg}, which can include futurely the new generation of GWs observatories already near completion\cite{thorne95}, the appropriate timing of both signals ($\nu$ + GWs) will provide the time-of-flight lag in between, i. e., the neutrino time delay will directly be stablished by both the observations\cite{MCDG00}, provided the source pinpointing by both capabilities be settled. 

Thus, for a 10 kpc distance, e. g., to the galactic center; for instance, the expected neutrino time lag should be: $ \Delta T_{GWs \longleftrightarrow \nu_s} = 0.515 $s, for a ($\nu_e$) neutrino mass $\leq 10 $eV, and energy $\leq 10$MeV, as in SN1987A. Thus the comparison between measured and theoretical time-of-flight delay will lead to a highly accurate estimate of the GWs velocity. An inferred mismatch between both timescales (expected and measured) may signal that the GWs speed as predicted by Einstein theory is not the correct one. This fact would positively point towards the NDL prediction as a more plausible explanation, since alternative theories as scalar-tensor gravity or other bi-metric gravitational theories predict that GWs travel at the speed of light, too.


\section{Birefringence of Gravitational Waves: The Method of the Effective Geometry}\label{geometry}

Just for later comparison we resume briefly the propagation of
electromagnetic waves in a non-linear regime. As it was shown
\cite{novello00} the non-linear photon propagates in a modified Riemannian
geometry characterized by an effective metric $g^{eff}_{\mu\nu}$ which 
is not the background one\footnote{Although the use of this formulation is not
mandatory, it simplifies greatly the analysis of the properties of
the wave propagation. Besides, we can describe the non-linear photon propagation
in a frame in which the electromagnetic 
forces on the photon are eliminated.} 

\be
g^{\mu\nu} = L_F\gamma^{\mu\nu} - 4\left[\left(
L_{FF} + \Omega L_{FG}\right)F^{\mu}\mbox{}_{\lambda}
F^{\lambda\nu} + \left(L_{FG} + \Omega L_{GG}\right)
F^{\mu}\mbox{}_{\lambda}F^{*\,\lambda\nu}\right].
\label{geral}
\ee

Here the scalar $\Omega$ obeys the equation: $\Omega^{2}\, \Omega_{1} + \Omega\,\Omega_{2} + \Omega_{3} = 0\label{r22}$. The proof of this was presented in \cite{novello00}.

\subsection{The Case of Gravitational Waves: \underline{One-parameter Lagrangians}}

Our main purpose in this section is to investigate the effects of 
nonlinearities in the equation of evolution of gravitational 
waves.  We will restrict the analysis in this section to the 
simple class of Lagrangians\footnote{The NDL theory is contained 
in this class.} defined by $L = L(U).$

From the least action principle we obtain the field equation 

\begin{equation}
\left[L_{U} F^{\lambda (\mu\nu)}\right]_{;\lambda} = 0.
\label{opl1}
\end{equation}

Using the antisymmteric and cyclic properties of the tensor $F_{\alpha,\beta\gamma}$  (and its covariant derivatives) for the discontinuity of the field equation (\ref{opl1}) through the Hadamard's surface $\Sigma$ we obtain

\begin{equation}
f_{\mu\,(\alpha\beta)} \,k^{\mu} + 2\,\frac{L_{UU}}{L_{U}}\,\xi\, 
F_{\mu\,(\alpha\beta)} \,k^{\mu} = 0
\label{opl2}
\end{equation} 

where $\xi$ is defined by $\xi  \doteq  F^{\alpha\beta\mu} \, f_{\alpha\beta\mu} - F^{\mu}\,f_{\mu}$. The consequence of such discontinuity in the identity (the analogous of the electromagnetic cyclic condition $F_{\{\mu \nu; \lambda\}} = 0$)

\be
F_{\alpha \beta}{^\nu_{\; ;\lambda}} + F_{\beta \lambda}{^\nu_{\; ;\alpha}} + F_{\lambda \alpha}^{\nu_{\; ;\beta}} = \frac{1}{2} \{ \delta^\nu_\alpha W_{[\beta \lambda]} +  \delta^\nu_\beta W_{[\lambda \alpha]} + \delta^\nu_\lambda W_{[\alpha \beta]} \},
\ee

with $W_{\alpha \beta} \doteq F_\alpha{\; ^\varepsilon}{_{\beta{;\varepsilon}}} - F_{\alpha, \beta}$, $W_{[\alpha \beta]} = W_{\alpha \beta} - W_{\beta \alpha}$, yields after some algebraic steps $\xi \eta_{\mu\nu}\, k^{\mu}\, k^{\nu} - 2\,F^{\alpha\mu\nu} \, 
f_{\beta\mu\nu}\, k^{\alpha}\,k^{\beta}  
+ F^{\alpha\beta\lambda} \, 
f_{\alpha}\, k_{\beta}\,k_{\lambda} +
F_{\alpha}\,f_{\beta}\,k^{\alpha}\,k^{\beta} = 0.
\label{gw4}$ From these equations we obtain the propagation equation for the field discontinuities

\be
L_{U} \eta^{\mu\nu} k_{\mu} k_{\nu} + 4L_{UU} (F^{\mu\alpha\beta} \, 
F^{\nu}\mbox{}_{\alpha\beta} -  \frac{1}{4} F^{\alpha\beta\mu} \, F_{\alpha\beta}\mbox{}^{\nu} - \frac{1}{2} F^{\mu} F^{\nu}) k_{\mu} k_{\nu} = 0.
\label{gww4}
\ee

Expression (\ref{gww4}) suggests that one can interpret the 
self-interaction of the background field $F^{\mu\nu\alpha},$ in what concerns the propagation of the discontinuities, 
as if it had induced a modification on the spacetime metric
$\eta_{\mu\nu}$, leading to the effective geometry

\be
g^{\mu\nu}_{\rm eff} = L_{U}\,\eta^{\mu\nu}  + 
4\, L_{UU}   ( F^{\mu\alpha\beta}\, F^{\nu}{}_{\alpha\beta} 
- \frac{1}{4}F^{\alpha\beta\mu}\, F_{\alpha\beta}\mbox{}^{\nu} - \frac{1}{2} F^{\mu} F^{\nu}).\label{geffec}
\ee

A simple inspection of this equation shows that in 
the particular case of the linear theory 
the discontinuity of the  gravitational field 
propagates along null paths in the Minkowski background. The more general case with two-parameter will be discussed elsewhere\cite{MVH00}. Thus the last equation confirms that the {\it propagation of gravitational waves also exhibit the birefringence phenomenon}.

\section{Astrophysical Test for Gravitational Waves Birefringence}

The MACHO Collaboration has announced recently that astronomical observations of starfields in our galaxy, using the Hubble Space Telescope (HST) and ground based telescopes, have provided compelling evidence for the existence of stellar-mass ($\sim 6$ M$_\odot$) black holes (BHs) adrift among the stars comprising the Milky Way\cite{bennett}. The two observations (1996 and 1998) revealed a subtle brightning of a background star produced by the microlensing gravitational enhancement of the light it emits due to the passage of an invisible object in between the star and Earth. A detalied analysis of the data ruled out white dwarfs or neutron stars as the lensing invisible source, and strongly  points towards dark stellar-mass objects (i. e., black holes) as the magnification sources since ordinary (massive) stars would be so bright to outshine the background star. These observations could have been supplemented by the discovery of multiple images of the lensed star but unfortunately the HST angular resolution is about two orders of magnitude larger than the minimum required for resolving (observing the separation of) a pair of images from it induced by the BH bending angle\cite{weinberg}

\be
{\hat\alpha}(r_0) =  2 \int^\infty_{r_0} \frac{e^{\mu/2} dr}{ \left[{\frac{r^4}{b^2}}e^{-\nu} -r^2 \right]^{1/2} } - \pi.
\ee
 
Here $r_0$ is the passage distance from the lensing object, $\mu$ and $\nu$ are the metric fields of a static spherically symmetric object, and $b$ is the impact parameter

\be
b = r_0 e^{-\nu(r_0)/2}.
\ee

In addition to these effects the starlight (radio waves, for instance) should undergo a time-delay respect to a pulse traveling in a region free of gravitation which may be measured by precise timing or throughout polarization patterns from the star. This effect, the Shapiro-Delay, is due to the light travel through changing gravitational fields. It was also predicted to exist for the case of binary radio pulsars.\cite{damour98,kopeikin99}

In the lines of this microlensing effect of starlight by a BH, analogously a gravitational wave (GW) signal from a galactic background source, a compact binary pulsar like PSR J1141-6545 (5 hours period), PSR 1534+12 or PSR 1913+16 should also be lensed (splitted) when passing near a massive compact dark object such as the MACHO Collaboration  BHs. Since both theoretical accurate estimates\cite{kalogera99} and observational statistical inferences\cite{camilo-group} of the abundance of galactic neutron star-neutron star binaries and coalescence rates of them are more promissing than earlier calculations, the following astronomical configuration looks a target to search for. Let us think for a while that a   galactic but distant binary radio pulsar is aligned with the lensing object (a Schwarszchild BH) and the Earth. A GW pulse is emitted from the binary, passes by the lens and is detected at Earth. Then according to general relativity both the polarization modes $h_+$ and $h_\times$ of the (linearized)  GW signal will undergo  deflection and time delay when flying-by the lens as in the case for light waves, but both will arrive to the detector at the same time, that is, there will be no time lag because in GR GWs travel at the speed of light and there is no birefringence effects on their propagation. Nevertheless, a phase lag for them in GR is predicted to be exactly $\pi/4$ radians.  This dephasing is expected to be measured by the new GWs detectors\cite{odylio}. As expected the signal power should be enhanced (enlargening of the GW amplitude) in a forseeable manner (For a more extensive discussion of this issue the reader is addressed to De Paolis, Ingrosso and Nucita (2000)\cite{depaolis00}).

Notwithstanding, in the NDL theory of gravity the existence of birefringence of the gravitational waves as described above will induce not only a rather different time delay but a phase lag too in the arriving GW signals, due to the different velocity of propagation $v_k$ for different spatial directions, as showed earlier (see Ref.\cite{NDL} for further details and definitions). This property may be tested with data collected with the forthcoming generation of GWs observatories such as LIGO, VIRGO, GEO-600, TIGAs, etc. cross-correlated with data from neutrinos, gamma-ray bursts and cosmic rays detectors\cite{herman00}. We have shown above that each polarization mode of the GW in the NDL theory is velocity-dependent (upon direction and magnitude). Then, the radiation component traveling at the lens left-hand side (from our point of view) will be affected in a different way compared with the right-handed component due to this global birefringence dependence. Thus the detected signals will be accordingly time lagged and phase-modulated in a way not mimicking GR, and such effects may be measured futurely. The above astrophysical scenario also works for a gravitational radiation source at the other side of our galaxy intervened by the Milky Way central black hole candidate Sagitarius A$^*$. Moreover, if the lens BH is a Kerr type one then the {\it frame dragging} (Lense-Thirring effect) induced by the BH spin would dramatically accentuate these effects, and it turns out that its observational verification will be a reachable endeavour in the days to come.


\vskip 1.0 truecm
 
\section{CONCLUSIONS}

Above we have shown how the almost simultaneous emission of GWs, GRBs and $\nu$s in a single astrophysical event may provide the non-gravitational processes that may turn the discrimination between general relativity and the NDL theory of gravity a reachable task in the near future. Prospective timing (detection) of such bursts from a unique source on the sky may prove powerful to settle the discrepancy between both theories in what concerns to the velocity of propagation of GWs. In this sense, the new generation of gravitational-wave observatories such as LIGO, VIRGO and GEO-600, together with the SNEWS neutrino network and the GRBs new detectors, and potentially the ultra high energy cosmic rays observatory AUGER, may prove useful. Moreover, because the neutrino energy can be measured by the time it gets the neutrino telescope and the source distance can be reliably estimated as discussed above, then from Eq.(\ref{neut}) the mass of the neutrino responsible for the observed event will be determined or stringently constrained by means not explored earlier. This will yield an innovative manner to check the threshold set to the neutrino mass by SuperKamiokande neutrino detector contained events. The point here is that despite the occurrence of several uncertainties (approximations, etc,) in the derivation of Eq.(\ref{neut}), the actual detection of both signals by the respectives observatories will render the task of constraining the velocitiy of propagation of the GWs and the mass of the neutrino involved in the process a feasible one. This is a paramount and inedit manner of weighting the neutrinos and measuring the speed of the gravitational waves.

\section*{Acknowledgments}

HJMC Acknowledges support from CLAF (Rio de Janeiro) and CNPq (Brazil) and the Abdus Salam ICTP (Trieste, Italy).

 
%

{\footnotesize

\end{document}